\documentclass[prd,twocolumn,amsmath,amsfonts,amssymb,showpacs]{revtex4} 
\usepackage{graphicx}
\usepackage{mathptmx}      
\usepackage{latexsym}
\def\beq{\begin{equation}}
\def\eeq{\end{equation}}
\def\bea{\begin{eqnarray}}

\def\eea{\end{eqnarray}}

\begin{document}

\title{ Phase transition, critical behavior, and critical exponents of Myers-Perry black holes}
\author{Mohammad Bagher Jahani Poshteh$^\ddag$, Behrouz Mirza$^{\ddag,\S}$}
\email{b.mirza@cc.iut.ac.ir}
\affiliation{$^\ddag$Department of Physics, Isfahan University of Technology, Isfahan, 84156-83111, Iran\\
 $^\S$Research Institute for Astronomy and Astrophysics of Maragha (RIAAM) -Maragha, P. O. Box 55134-441, Iran}

\author{Zeinab Sherkatghanad$^{\ddag}$}
\email{ z.sherkat@ph.iut.ac.ir}
\affiliation{$^\ddag$Department of Physics, Isfahan University of Technology, Isfahan, 84156-83111, Iran\\
}



\begin{abstract}
The critical behavior of Myers-Perry black holes with equal angular momenta in even dimensions are studied. We include the corrections beyond the semiclassical approximation on Hawking temperature in the grand canonical ensemble. Having done so, we find that the critical behavior and critical exponents of Myers-Perry black holes correspond to those of a Van der Waals liquid-gas where this analogy holds in any dimension. Also, using  Ehrenfest's equations, we calculate the order of the phase transition in the semiclassical approximation for the canonical ensemble and beyond the semiclassical approximation for the grand canonical ensemble near the critical point. Finally, the Ruppeiner curvature formula is used to investigate the thermodynamic geometry of Myers-Perry black holes.
\end{abstract}
\maketitle

\section{INTRODUCTION}
The Hawking  temperature of a black hole is proportional to
its surface gravity. Also a black hole's entropy is proportional to its horizon area. This is known as
the celebrated Bekenstein-Hawking area law $S_{BH} =\frac{A}{4}$.
There are several different methods for calculating the
corrections to the semiclassical Bekenstein-Hawking entropy. These
are based on statistical mechanical arguments, field theory
methods, quantum geometry, the Cardy formula, the generalized uncertainty principle, etc.
(The corresponding literature is rather extensive; for a partial
selection, see Refs. \cite{c1,c2,c3,c4,Haw,RG,GUP,GUP1}.)\\
\indent One of the black hole solutions in higher
dimensions which has attracted  a lot of attention is the  Myers-Perry black hole \cite{aman1,aman2,aman3,aman4,aman5,aman6}, whose uncharged
rotating version is a direct generalization of the Kerr black hole
solution in General Relativity. The classification of black
hole species in higher dimensions was studied by
Rodriguez~\cite{esp}. Also, the corrections beyond the semiclassical approximation can lead  to  corrected Hawking temperature and entropy for Myers-Perry (MP) black holes.\\
\indent The lack of a statistical description for black hole
systems has encouraged researchers to consider thermodynamic geometry \cite{R1,R2a,R2b,R2c,R3,R4,momen1}, Ehrenfest's equations \cite{E1,E2,E3}, and, recently, the analogy of a black hole with a  Van der Waals system \cite{wu1,m1,m2,m3}. An important part
in the theory of phase transitions is the exploration of the thermodynamic behavior
of a system near its critical point using  critical exponents. These critical exponents are
supposed to be universal and are independent of the details of the interaction.
In other words, different physical systems may share the same critical exponents \cite{prig,crit1,crit2,crit3}. In this way, we may find the possibility of searching for the nature of phase transitions and critical exponents
in the grand canonical ensemble (for which the angular velocity, $\Omega$, is taken to be fixed) and in the canonical ensemble (for which the angular momentum, $J$, is also taken to be fixed). We consider the corrected temperature of MP black holes beyond the semiclassical approximation in the grand canonical ensemble and study the analogy of this system with a liquid-gas system. We also find that the critical exponents  correspond to a Van der Waals system in the grand canonical ensemble. For both the canonical and grand canonical ensembles we search for the satisfication of Ehrenfest's equations at the critical point \cite{E4,E5,E6}.\\
\indent The outline of this paper is as follows. In Sec. II, we
consider the semiclassical thermodynamic quantities and critical behavior of MP
black holes in even dimensions with $n$ nonzero equal spins $J$ in the grand canonical and canonical ensembles. In
Sec. III, corrections to the semiclassical Hawking temperature
and entropy are investigated in the grand canonical ensemble and the analogous behavior of this system to a Van der Waals gas is studied. In
Sec. IV, the Ehrenfest's equations for these black hole systems are developed in both ensembles and the order
of the phase transition near the critical points is analyzed.
In Sec. V, the critical exponents near the critical point
are calculated in the grand canonical ensemble. Section VI is devoted to the  study of the thermodynamic geometry of MP black holes and it is found that the scalar curvature diverges exactly at the point where the  heat capacity is divergent.
\section{SEMICLASSICAL THERMODYNAMICS OF MYERS-PERRY BLACK HOLES}
\begin{figure}[h]
\centering
\includegraphics[angle=0,width=7cm,keepaspectratio]{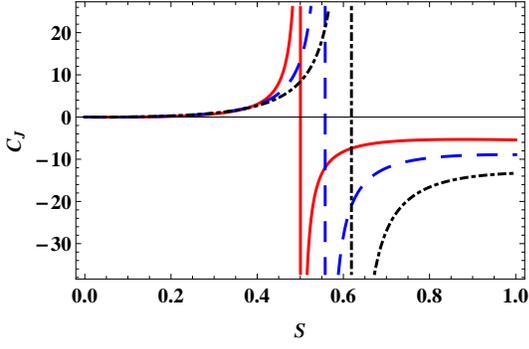}
\caption[]{{Specific heat $C_{J}$ with respect to $S$ for
$d=6 \ [red\ (solid)\ line], \ d=8\ [blue\ (dashed)\ line],\  d=10\
[black\ (dashed-dotted)\ line]$ and $J=0.1$.}} \label{figure 1}
\end{figure}

A summary of the thermodynamic quantities
of MP black holes in even dimensions with $n=\frac{d-2}{2}$ nonzero equal
spins $J$ is presented below. More details and a complete solution can be found in Ref. \cite{aman3}.
The multiple-spin Kerr black hole's metric
in Boyer-Lindquist coordinates for an even $d$ is given by \cite{aman4,aman5}
     \bea
    \label{a1}
    &&ds^2 = -dt^{2} +r^{2} d \alpha ^{2}+ (r^{2}+ a ^{2}_{i})(d\mu ^{2}_{i} + \mu ^{2}_{i} d\phi ^{2}_{i})\\\nonumber
    &&+\frac{m r}{\Pi F}(dt - a_{i} \mu^{2}_{i} d\phi_{i})^{2} + \frac{\Pi F}{\Pi - m r}dr^{2},
  \eea
where  $\mu ^{2}_{i}+\alpha ^2=1$ and $m=\frac{16\pi G M}{(d-2)\Omega_{(d-2)}}$. The functions $\Pi$ and $F$ are defined as
    \bea
     \label{a2}
     &&\Pi = \prod_{i=1}^{(d-1)/2} (r^{2}+a^{2}_{i}),\\
     &&F=1-\frac{a_{i}^{2}\mu_{i}^{2}}{r^{2} + a^{2}_{i}}.
      \eea
The metric is slightly modified for an odd $d$. The event horizon in the Boyer-Lindquist coordinates for an even $d$ is defined by
     \bea
     \label{a3} \Pi(r_+) -m r_+ = 0.
     \eea
Moreover, the area of the event horizon for an even $d$ can be expressed as follows:
     \bea
      A=\Omega_{(d-2)} \prod_{i} (r^{2}_{+} + a^{2}_{i}).
     \eea

\begin{figure}[h]
\centering
\includegraphics[angle=0,width=7cm,keepaspectratio]{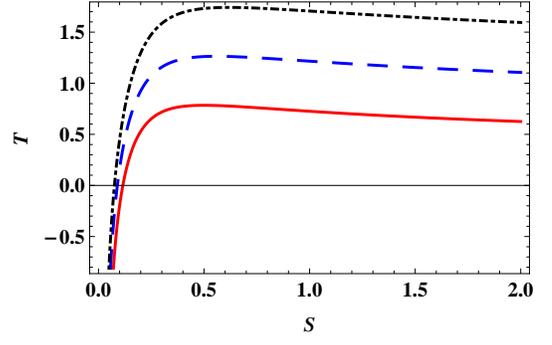}
\caption[]{{Semi-classical Hawking temperature, $T$, with
respect to $S$ for $d=6 \ [red\ (solid)\ line],\  d=8\ [blue\ (dashed)\ line],\  d=10  \
[black\ (dashed-dotted)\ line]$ and $J=0.1$.}} \label{figure 2}
\end{figure}

\begin{figure}[h]
\centering
\includegraphics[angle=0,width=7cm,keepaspectratio]{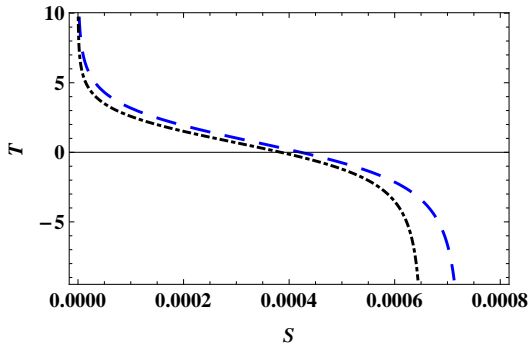}
\caption[]{{Semi-classical Hawking temperature, $T$, with
respect to $S$ for $ d=8\ [blue\ (dashed)\ line],\   d=10\
[black\ (dashed-dotted)\ line]$ and $\Omega=10$.}} \label{figure 3}
\end{figure}

\begin{figure}[h]
\centering
\includegraphics[angle=0,width=7cm,keepaspectratio]{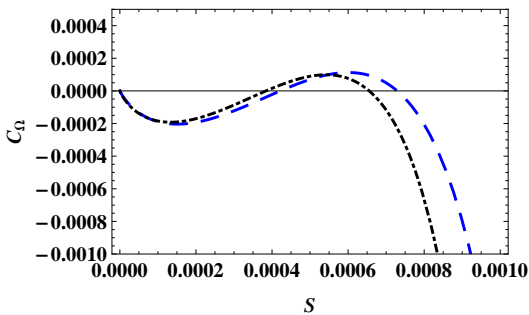}
\caption[]{{Specific heat $C_{\Omega}$ with respect to $S$ for
$d=8\ [blue\ (dashed)\ line], \  d=10\
[black\ (dashed-dotted)\ line]$ and $\Omega=10$.}} \label{figure 4}
\end{figure}

The Bekenstein-Hawking entropy for the MP black holes with an even $d$ is given by
        \bea
        S=\prod_{i} (r^{2}_{+} + a^{2}_{i}),
        \eea
where $k_B = {1 \over \pi}$ and $G =\frac{\Omega_{(d-2)}}{4\pi}$.
Briefly put, the mass and Hawking temperature of the black hole are
   \bea
   M=\frac{d-2}{4}
   S^{(\frac{d-3}{d-2})}\sqrt{1+\frac{4J^{2}}{S^{2}}}
   \eea
and
  \bea
   \label{a4}
   &&T= \frac{(d-3)}{4 S^{\frac{1}{d-2}}}\frac{1-\frac{4J^{2}}{(d-3)S^{2}}}{\sqrt{1+\frac{4J^{2}}{S^{2}}}}.
  \eea
The angular velocity is defined as
  \bea
  \label{a5}\Omega = (\frac{\partial M}{\partial J})_{S} = \frac{(d-2)S^{\frac{d-3}{d-2}} J}{\sqrt{1+\frac{4J^{2}}{S^{2}}} S^{2}}.
  \eea
Thus, the heat capacity for MP black holes in $d$ dimensions with $n=\frac{d-2}{2}$ nonzero equal spins in the canonical (constant $J$) ensemble is defined by
   \bea
   \label{a6}
   &&C_{J}(J, S)=T (\frac{dS}{dT})_{J}\\\nonumber
    &&=\frac{(4J^{2}+S^{2})((d-3)S^{2}-4J^{2})(d-2)S}{16(d -1) J^4+4(6 +(d-4)d) J^2 S^2 - (d-3) S^4}.
   \eea
\noindent Now, we can see that the heat capacity for the canonical ensemble in Eq. (\ref{a6}) is divergent at the critical entropy which is given by
 \bea
 \label{a8}
  S_{c}&=&\frac{J}{\sqrt{d-3}} (2(-4d+6+d^2+\\\nonumber
  && \sqrt{32d^2-64d-8d^3+48+d^4}))^\frac{1}{2}.
  \eea
The discontinuity in the plot of the heat capacity indicates that a kind of phase transition occurs at this point. (Fig. \ref{figure 1}). Also, the Hawking temperature has  positive values at this critical point $S=S_c$ (Fig. \ref{figure 2}).  In order to find the nature of this phase transition, it is necessary to consider  Ehrenfest's equations near the critical point. This phase transition will be dealt with in greater detail below.\\
The heat capacity in the grand canonical  ensemble (constant $\Omega$) is defined as
   \bea
   \label{a7}
   &&C_{\Omega}(\Omega, S)=T (\frac{dS}{dT})_{\Omega}\\\nonumber
     &&=\frac{-S}{-d^{3}+7 d^{2}-16 d+12+4 d \Omega^{2} S^{\frac{2}{d-2}} -16 \Omega^{2} S^{\frac{2}{d-2}}}\\\nonumber
     &&\times(8S^{\frac{2}{d-2}} d^{2} \Omega^{2} -d^{4}+9 d^{3}-30 d^{2}-36 S^{\frac{2}{d-2}} d \Omega^{2}\\\nonumber
     &&+44 d+40 S^{\frac{2}{d-2}}\Omega^{2}  - 24-16 S^{\frac{4}{d-2}}\Omega^{4}).
     \eea
\indent Based on Eqs. (\ref{a4}), (\ref{a5}), and (\ref{a7}), we can plot the temperature and heat capacity with respect to $S$ in the grand canonical ensemble (Figs. \ref{figure 3} and  \ref{figure 4}). We find that $T$ is equal to zero at a special point whose value depends on the dimensions $d$ and angular velocity $\Omega$ (for $d=8$, $S=0.000421875$).
Since the Hawking temperature, $T$, has  positive values before these specific points and the heat capacity is negative over the entire region, the black hole system is unstable throughout. The heat capacity is continuous for all values of $d$ and $\Omega$; therefore, the existence of a phase transition is ruled out.
\section{ANALYSIS OF PHASE TRANSITION BEYOND THE SEMICLASSICAL APPROXIMATION}
In the previous section, we showed that the heat capacity in the grand canonical ensemble (fixed $\Omega$) is not divergent. Recently, a phase transition has been investigated for the corrected thermodynamics of a Kerr black hole $(d=4)$ beyond the semiclassical approximation in Ref. \cite{E3}. In this section, we would like to explore these calculations for the arbitrary dimensions, in this case for MP black holes. We also study the analogy between this system and a liquid-gas system.\\
We consider the corrections beyond the semiclassical approximation to the  Bekenstein-Hawking entropy of black holes and the Hawking temperature which can be obtained by using a variety of approaches,
based on statistical mechanical arguments, field theory methods, quantum geometry, the Cardy formula, the generalized uncertainty principle, etc. A review of these methods may be found in Refs. \cite{c1,c2,c3,c4,cardy,Haw,RG,GUP,GUP1}.
The first-order correction to the entropy can be expressed by
     \bea
     \label{c2}  \tilde{S}=S+\beta_{1}\frac{4}{(d-2)}\log{S}.
     \eea
The first term is the semiclassical entropy $S=\frac{4M r_{+}}{\hbar(d-2)}$ for MP black holes and the second term is the first-order quantum correction. Also, $\beta_{1}$ is a dimensionless constant and smaller than $1$. To identify the coefficients of the leading corrections such as $\beta_{1}$, we can use the trace anomaly or other standard methods.\\
Hawking in Ref. \cite{Haw} calculated one-loop corrections to the radiation process, associated with a trace anomaly and showed that the backreaction has an interesting effect on the Hawking radiation and temperature. Considering the results of the renormalization group approach  \cite{RG} and the generalized uncertainty principle \cite{GUP,GUP1} suggests  that in the simplest cases the first-order correction term to the Hawking temperature is proportional to the inverse of the area (see also Ref. \cite{c1}).  The above-mentioned theories suggest the following first-order correction to the Hawking temperature, which is also consistent with the first law of thermodynamics:

     \bea
    \label{c3} \tilde{T}= T(1- \beta_{1} \frac{4}{(d-2)S}).
     \eea
It should be noted that a similar first-order quantum correction was also calculated in Ref. \cite{B1}; however,  their assumptions have no theoretical justification and were criticized in Ref. \cite{Yale}. \\
\begin{figure}[h]
\centering
\includegraphics[angle=0,width=7cm,keepaspectratio]{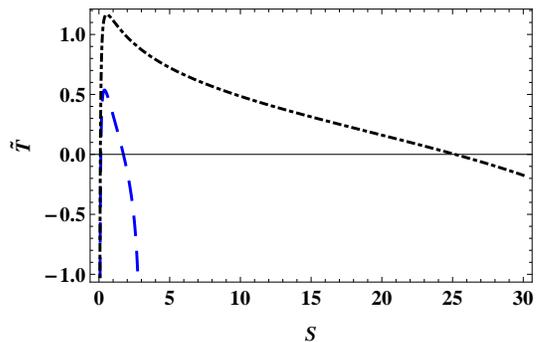}
\caption[]{{Corrected Hawking temperature $\tilde {T}$ with
respect to $S$ for $\Omega< \Omega_{c}$, $\beta_{1}=0.2$ and $ d=8\ [blue\ (dashed)\ line],\  d=10\
[black\ (dashed-dotted)\ line]$.}}
\label{figure 5}
\end{figure}
\begin{figure}[h]
\centering
\includegraphics[angle=0,width=7cm,keepaspectratio]{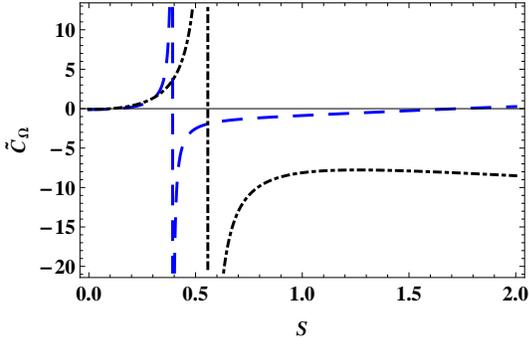}
\caption[]{{Corrected specific heat $\tilde {C}_{\Omega}$
with respect to $S$ for $\Omega< \Omega_{c}$, $\beta_{1}=0.2$ and $d=8\ [blue\ (dashed)\ line],\  d=10\
[black\ (dashed-dotted)\ line]$.}} \label{figure 6}
\end{figure}

\begin{figure}[h]
\centering
\includegraphics[angle=0,width=7cm,keepaspectratio]{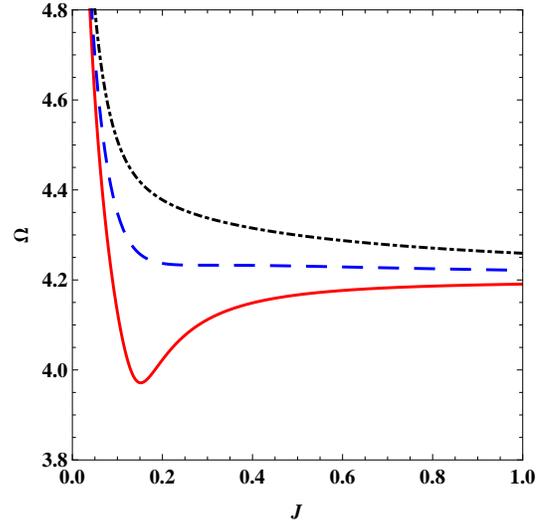}
\caption[]{{The angular villosity $\Omega$ with respect to the angular momentum $J$ at $\tilde{T}=\tilde{T}_c \ [blue\ (dashed)\ line]$, $\tilde{T}> \tilde{T}_c\
[black\ (dashed-dotted)\ line]$ and $\tilde{T}< \tilde{T}_c  \ [red\ (solid)\ line]$ for $d=6$ and $\beta_{1}=0.2$.}} \label{figure 7}
\end{figure}

\begin{figure}[h]
\centering
\includegraphics[angle=0,width=7cm,keepaspectratio]{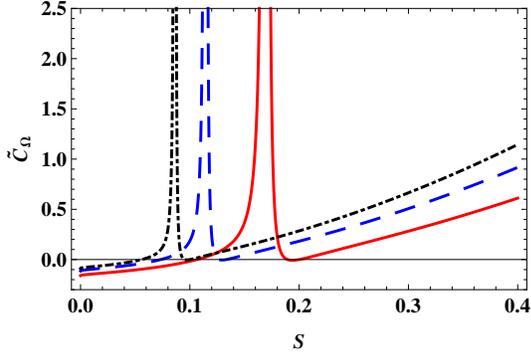}
\caption[]{{Corrected specific heat $\tilde {C}_{\Omega}$
with respect to $S$ for $\beta_{1}=0.2$, $\Omega=\Omega_c$ and $d=6 \ [red\ (solid)\ line],\  d=8\ [blue\ (dashed)\ line],\  d=10\
[black\ (dashed-dotted)\ line]$.}} \label{figure 8}
\end{figure}

\noindent Equations (\ref{a4}), (\ref{a5}), (\ref{c2}) and (\ref{c3}) can now be used to calculate the corrected temperature and
specific heat up to the first-order correction as a function of entropy and angular velocity for these types of black holes,
      \bea
      \label{c33}
      && \tilde{T}(S,\Omega)=(1-\beta_{1} \frac{4}{(d-2) S})\times\\\nonumber
      && \frac{(d-3) (1-\frac{4S^{2} \Omega^{2}}{(d-3) (-4 S^{2} \Omega^{2}+S^{\frac{2(d-3)}{d-2}}(d-2)^{2})})}
     {4 S^{\frac{1}{d-2}} \sqrt{1+\frac{4S^{2} \Omega^{2}}{-4 S^{2} \Omega^{2}+S^{\frac{2(d-3)}{d-2}}(d-2)^{2}}}},
     \eea
and
     \bea
     \label{c4}
     \tilde{C}_{\Omega}(S,\Omega)&=&\tilde{T}(\frac{d\tilde{S}}{d\tilde{T}})_{\Omega}\\\nonumber
     &=&\frac{ ((d-2)^2 S^2 - 16 \beta _1 ^2)}{\zeta} ((d-3) (d-2)^3 \\\nonumber
     &-& 4 (d-2) (2 d-5) S^{\frac{2}{d-2}} \Omega ^2 +16 S^{\frac{4}{d-2}} \Omega ^ 4),
     \eea
where
   \bea
     \label{c5}
     \zeta &=&(d-2) ((3-d) (d-2)^3 S\\\nonumber
      &+& 4 (d-4) (d-2) S^{\frac{d}{d-2}} \Omega ^2\\\nonumber
     &+&4 ((d-3 ) (d-2 )^2 (d-1)\\\nonumber
     &-&8 (d-3) (d-1) S^{\frac{2}{d-2}}\Omega ^2 \\\nonumber
     &+&16 S^{\frac{4}{d-2}} \Omega ^4) \beta _1 )-2 \Omega ^2+16 S^{\frac{4}{d-2}} \Omega ^4.
     \eea
Although the corrected specific heat $\tilde{C}_{\Omega}$ is divergent at two specific points $S_{(1,2)}$ for a given dimension as $d$ indicated in Eq. (\ref{c5}), the corrected Hawking temperature in Eq. (\ref{c33}) does not have a real value at the larger point $S_2$. This means that the larger divergent point $S_2$ is nonphysical (Figs. \ref{figure 5} and \ref{figure 6}). Based on Eqs. (\ref{a5}) and (\ref{c3}), the angular velocity $\Omega$ can be depicted with respect to the angular momentum $J$ as a "P-V diagram" at $\tilde{T}=\tilde{T}_c$, $\tilde{T}>\tilde{T}_c$ and $\tilde{T}< \tilde{T}_c$ for the  given dimensions $d$ (Fig. \ref{figure 7}). An inflection point can be observed at $\tilde{T}<\tilde{T}_c$; this behavior is similar to that of a Van der Waals system. The critical point is obtained from $(\frac{\partial \Omega}{\partial J})_c=0$ and $(\frac{\partial ^2 \Omega}{\partial J ^2})_c=0$. Thus, by using the equation of state in  higher dimensions $d$ ($\Omega$ as a function of $\tilde{T}$, $d$,  and $J$) and also $(\frac{\partial \Omega}{\partial J})_c=0$ and $(\frac{\partial ^2 \Omega}{\partial J ^2})_c=0$, we can obtain the values of $\Omega_c$, $S_c$, and $\tilde{T}_c$ at the critical point when the  discriminant of Eq. (\ref{c5}) vanishes, in other words, when two divergent points meet at one critical point (Fig. \ref{figure 8}).\\
Stability is determined by the third derivative of $\Omega$ with respect to $J$. The inequality $(\frac{\partial ^3 \Omega}{\partial J ^3})_c< 0$ \cite{prig} shows the stability at the critical point. The results are summarized in Table $I$.\\
\indent The critical values of $\Omega_c$, $S_c$, and $\tilde{T}_c$ depend on the coefficient $\beta_1$. This coefficient  is not exactly known but is smaller than one. However, for $d=4$, and for  $\frac{P_c V_c}{T_c}=\frac{3}{8}$ which is universal and exactly the same as a Van der Waals fluid, the value for this coefficient will be $\beta_{1}^{d=4}=\frac{1}{24\sqrt{3}}$. Alternatively, its value for a given $d$ may be determined by considering the relation $\frac{P_c V_c}{T_c}=\frac{2d-5}{4d-8}$ \cite{m2}; thus,  $\beta_{1}^{d=6}= \frac{7 (5-5^{\frac{1}{3}})}{1200}=0.01919$ and $\beta_{1}^{d=8}=0.01820$. It is shown below that  we may obtain the ratification of both of Ehrenfest's equations and the critical exponents at the critical point in the canonical and grand canonical ensemble.
\begin{table}[h]
\centering
\begin{tabular}{|c|c|c|c|c|}
\hline \cline{2-3}
 $d$ & $\Omega_c$ & $S_c$ & $\tilde{T}_c$ & $(\frac{\partial ^3 \Omega}{\partial J ^3})_c$\\
\hline\hline
    6 & $3.03198$ & $0.1696$& $0.2371850$ &$ -25865.89$  \\ \hline
    8 & $4.23272$   & $0.11445$& $0.258412$ &$-7383.72$ \\ \hline
    10& $5.36657$   & $0.0863001$ &$0.277826$ &$-61407.8$ \\ \hline
\end{tabular} \caption{Summary of results for the values of $\Omega_c$, $S_c$, $\tilde{T}_c$ and $(\frac{\partial ^3 \Omega}{\partial J ^3})_c$ for $\beta_{1}=0.2$.}
\end{table}
\section{ EHRENFEST'S EQUATIONS AT THE CRITICAL POINT}
In this section, we study the satisfication of  Ehrenfest's equations at the critical point in the grand canonical and the canonical ensemble for the black hole system \cite{E3, E4, E5}.  Ehrenfest's equations for MP black holes in the grand canonical ensemble can be expressed as
      \bea
     &&-(\frac{\partial \Omega}{\partial \tilde{T}})_{\tilde{S}}=\frac{\tilde{C}_{\Omega_{2}}-\tilde{C}_{\Omega_{1}}}{\tilde{T}
     J(\tilde{\alpha}_{2}-\tilde{\alpha}_{1})},\\
      &&-(\frac{\partial \Omega}{\partial
      \tilde{T}})_{J}=\frac{\tilde{\alpha}_2-\tilde{\alpha}_1}{\kappa_{\tilde{T}_2}-\kappa_{\tilde{T}_1}},
     \eea
where
   \bea
     \tilde{\alpha}=\frac{1}{J} (\frac{\partial J}{\partial \tilde{T}})_{\Omega},\\
      \kappa_{\tilde{T}}=\frac{1}{J}(\frac{\partial J}{\partial
      \Omega})_{\tilde{T}}.
     \eea
A discontinuity in the plots of $\tilde{\alpha}$ and $\kappa_{\tilde{T}}$ is also necessary for a phase transition to take place. By using the chain rule of partial differentiation and the definitions of $\Omega$ and $\tilde{T}$ in Eqs. (\ref{a5}) and (\ref{c33}), respectively, we can plot $\tilde{\alpha}$ and $\kappa_{\tilde{T}}$ with respect to $S$. The results show that $\tilde{\alpha}$ and $\kappa_{\tilde{T}}$ are divergent exactly at the critical point.\\
\indent In what follows we present the  calculations  to identify the validity of Ehrenfest's equations and the order of the phase transition for MP black holes in the grand canonical ensemble. Consider $\tilde{C}_\Omega=\frac{f(S)}{g(S)}$, $J \tilde{\alpha}=\frac{h(S)}{g(S)}$ and $J \kappa_{\tilde{T}}=\frac{k(S)}{g(S)}$ such that for the function $g(S_c)=0$  near the critical point, we have
    \bea
     \tilde{C}_{\Omega_2}-\tilde{C}_{\Omega_1}&=&\frac{f(S_2)}{g(S_2)}-\frac{f(S_1)}{g(S_1)}\\\nonumber
     &=&f(S_c)(\frac{1}{g(S_{c_2})}-\frac{1}{g(S_{c_1})}),
     \eea
where  $\tilde{C}_{\Omega}\mid _{S_i}= \tilde{C}_{\Omega_i}$, and $S_2=S_c+\varepsilon$ and $S_1=S_c -\varepsilon$. Thus, the rhs of both Ehrenfest's equations are given by
     \bea
     &&\frac{\tilde{C}_{\Omega_ 2}-\tilde{C}_{\Omega_1}}{\tilde{T} J(\tilde{\alpha}_{2}-\tilde{\alpha}_{1})}=\frac{f(S_c)}{\tilde{T}_c h(S_c)},\\
     &&\frac{J (\tilde{\alpha}_{2}-\tilde{\alpha}_{1})}{J (\kappa_{\tilde{T}_2}-\kappa_{\tilde{T}_1})}=\frac{h(S_c)}{k(S_c)}.
     \eea
Both sides of the first Ehrenfest's equation are calculated for any values of  $d$ and $\beta _1$ at the critical point in the grand canonical ensemble,
\bea
    rhs=lhs&=&(1+\frac{4 \beta _1}{(d-2) S_c}) S_c^{ \frac{1}{d-2}}\\\nonumber
    &\times& ((d-2)^2 -4\Omega_c ^2 S_c^ {\frac{2} {d-2}})^{\frac{3}{2}}.
 \eea
It is also observed  that both sides of the second Ehrenfest's equations are equal. Ehrenfest's equations for the MP black holes in the canonical ensemble are given by
    \bea
     &&-(\frac{\partial J}{\partial T})_{S }=\frac{C _{J_{2}}-C _{J_{1}}}{T\Omega(\alpha_{2}-\alpha_{1})},\\
      &&-(\frac{\partial J}{\partial T})_{\Omega}=-\frac{\alpha_{2}-\alpha_{1}}{\kappa_{T_2}-\kappa_{T_1}},
    \eea
where
   \bea
    \alpha=-\frac{1}{\Omega} (\frac{\partial \Omega}{\partial T})_{J},\\
     \kappa_{T}=\frac{1}{\Omega}(\frac{\partial \Omega}{\partial J})_{T}.
     \eea
In this ensemble we can see that $\alpha$ and $\kappa_{T}$ are divergent exactly at the point where the specific heat is divergent (at $S_c$) and also that the first and second Ehrenfest's equations are satisfied at the critical point. So, a second-order phase transition is taking place in the grand canonical and canonical ensembles. \\
\section{GIBBS FREE ENERGY}
\begin{figure}[h]
\centering
\includegraphics[angle=0,width=7cm,keepaspectratio]{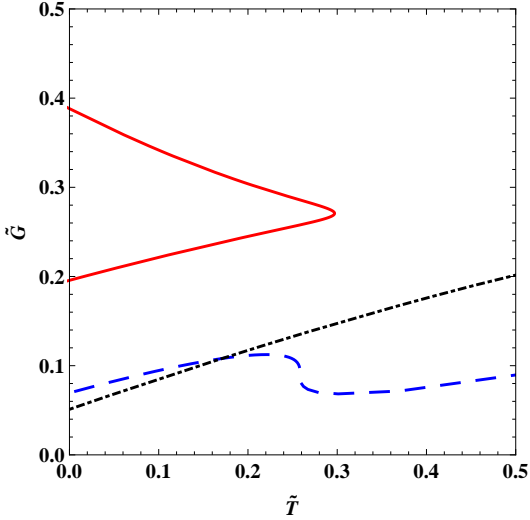}
\caption[]{{The corrected Gibbs free energy with respect to the corrected temperature for $\beta _1=0.2$ and
$\Omega < \Omega_{c} \ [red\ (solid)\ line],\  \Omega =\Omega_{c}\ [blue\ (dashed)\ line],\  \Omega> \Omega_{c}\
[black\ (dashed-dotted)\ line]$}.}
\label{figure 9}
\end{figure}

\begin{figure}[h]
\centering
\includegraphics[angle=0,width=7cm,keepaspectratio]{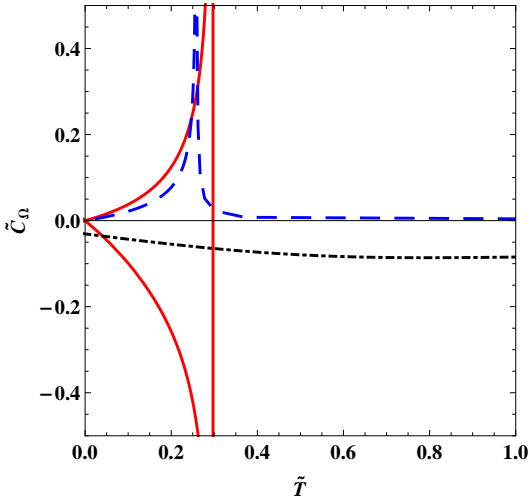}
\caption[]{{The corrected heat capacity with respect to the corrected temperature $\beta _1=0.2$ and
$\Omega<\Omega_{c} \ [red\ (solid)\ line],\  \Omega =\Omega_{c}\ [blue\ (dashed)\ line],\  \Omega>\Omega_{c}\
[black\ (dashed-dotted)\ line]$}.}
\label{figure 10}
\end{figure}

In this section, we consider the behavior of the corrected Gibbs free energy and the corrected specific heat as a function of temperature for different values of $\Omega$ in the grand canonical ensemble. The corrected Gibbs free energy for MP black holes can be described by Eqs. (\ref{a5}), (\ref{c2}), (\ref{c3}), and the following relation:
\bea
\label{G1}
\tilde{G}(d,\Omega ,\tilde{S})=M-\tilde{T} \tilde{S}-\Omega J.
\eea
The corrected specific heat in the grand canonical ensemble, however, is determined by  Eq. (\ref{c4}). Now we can plot the corrected Gibbs free energy and the corrected specific heat with respect to $\tilde{T}$ at $\Omega =\Omega_{c}$, $\Omega > \Omega_{c}$ and $\Omega < \Omega_{c}$ (Figs. \ref{figure 9} and \ref{figure 10}). Consider the $\tilde{G}-\tilde{T}$ plot; there are two wings at $\Omega < \Omega_{c}$, which are joined at the special point $\tilde{T}_1$ (Fig. \ref{figure 9} ). The corrected heat capacity is divergent and changes from negative values (unstable phase) to positive ones (stable phase) at $\tilde{T}_1$; the black hole system also exists just at $\tilde{T} < \tilde{T}_1$. Since the corrected Gibbs free energy for the upper wing does not have its minimum value, the upper wing behaves like  a metastable state. It is also observed  that the black holes experience a second-order phase transition at $\tilde{T}_1$ which is not in a metastable state, as shown in Fig. \ref{figure 9}. Clearly, the critical exponents are completely different from the Van der Waals system at this point. In other words, at this critical point we get $\frac{\partial ^2 \Omega}{\partial J ^2}\neq 0$ and this behavior leads to different values for the critical exponents. Thus, what have been left out are the critical exponents of this second-order phase transition.\\
For $\Omega=\Omega_{c}$, the corrected Gibbs free energy is positive for any value of  $\tilde{T}$ and the corrected specific heat is divergent at $\tilde{T}=\tilde{T}_c$. For $\Omega> \Omega_{c}$, the corrected specific heat is negative for all values of $\tilde{T}$ and is not divergent at any special point  (Fig. \ref{figure 10}). These calculations can be expanded for $d\geq 4$ dimensions, the results being thus independent of dimensions.

\section{CRITICAL EXPONENTS}
In this section, we calculate the critical exponents of MP black holes near the critical point in the grand canonical ensemble. It was shown in the previous section that not only do the black holes in the grand canonical ensemble behave similar to a van der Waals system at the critical point, but also that the corrected Gibbs free energy has positive values throughout. If we consider the entropy near the critical point $S_c$, where $\tilde{C}_\Omega$ is divergent, we get\cite{crit1,crit2}
\bea
\label{crit1}
S=S_c(1+\Delta),
\eea
where $\mid\Delta\mid\ll 1$. Since the corrected temperature is a function of the semiclassical entropy, we have
\bea
\label{crit2}
\check{T}=\tilde{T}_c(1+\epsilon).
\eea
Here $\mid\epsilon\mid\ll 1$. The critical exponent $\alpha $ is associated with the singular behavior of the corrected specific heat $\tilde{C}_{J} $. Since $\tilde{C}_J$ does not diverge at this critical point $S_c$ (where $\tilde{C}_\Omega$ is divergent), we find that the critical exponent $\alpha=0$. By expanding $\Omega(S,\tilde{T})$ and $J(S,\Omega)$ near the critical point, we get the following two equations:
\bea
\label{crit3}
&&\Omega (S,\tilde{T})=\Omega _c +[(\frac{\partial \Omega}{\partial S})]_{\tilde{T}=\tilde{T}_c, S=S_c} (S-S_c)\\\nonumber
&&+\frac{1}{2}[(\frac{\partial ^2 \Omega}{\partial S ^2})]_{\tilde{T}=\tilde{T}_c, S=S_c} (S-S_c)^2\\\nonumber
&&+\frac{1}{6} [(\frac{\partial ^3 \Omega}{\partial S ^3})]_{\tilde{T}=\tilde{T}_c, S=S_c} (S-S_c)^3\\\nonumber
&&+[(\frac{\partial \Omega}{\partial \tilde{T}})]_{\tilde{T}=\tilde{T}_c, S=S_c} (\tilde{T}-\tilde{T}_c)\\\nonumber
&&+[(\frac{\partial ^2 \Omega}{\partial S \partial \tilde{T}})]_{\tilde{T}=\tilde{T}_c, S=S_c} (\tilde{T}-\tilde{T}_c) (S-S_c),
\eea
and
\bea
\label{crit4}
&&J(S,\Omega)=J_c +[(\frac{\partial J}{\partial S})]_{\Omega =\Omega _c, S=S_c} (S-S_c)\\\nonumber
&&+higher\ order\ terms.
\eea
Since $(\frac{\partial \Omega}{\partial J})_c=0$ and $(\frac{\partial ^2 \Omega}{\partial J ^2})_c=0$
at the critical point, we can use the chain rule of partial
 differentiation at this point and rewrite Eq. (\ref{crit3}) in the following form:
\bea
\label{crit5}
&&\Omega(S,\tilde{T})=\Omega _c +\frac{1}{6} [(\frac{\partial ^3 \Omega}{\partial S ^3})]_{\tilde{T}=\tilde{T}_c, S=S_c} (S-S_c)^3\\\nonumber
&&+[(\frac{\partial \Omega}{\partial \tilde{T}})]_{\tilde{T}=\tilde{T}_c, S=S_c} (\tilde{T}-\tilde{T}_c)\\\nonumber
&&+[(\frac{\partial ^2 \Omega}{\partial S \partial \tilde{T}})]_{\tilde{T}=\tilde{T}_c, S=S_c} (\tilde{T}-\tilde{T}_c) (S-S_c).
\eea
Differentiating $\Omega(S,\tilde{T})$ for a fixed $\tilde{T}$ and using both Eq. (\ref{crit4}) and Maxwell's equal-area law \cite{m1}  yields the critical exponent $\beta=\frac{1}{2}$. Let us obtain the critical exponent $\gamma$ associated with $\kappa_{\tilde{T}}^{-1} \propto (\frac{\partial \Omega}{\partial J})_{\tilde{T}}$. $J(S,\tilde{T})$ is first expanded  near the the critical point,
\bea
\label{crit6}
&&J(S,\tilde{T})=J_c +[(\frac{\partial J}{\partial S})]_{\tilde{T}=\tilde{T}_c, S=S_c} (S-S_c)\\\nonumber
&&+[(\frac{\partial \Omega}{\partial \tilde{T}})]_{\tilde{T}=\tilde{T}_c, S=S_c} (\tilde{T}-\tilde{T}_c)+ \ {higher}\ order\ terms.
\eea
 Equations (\ref{crit5}) and (\ref{crit6}) are differentiated with respect to entropy to obtain
\bea
\label{crit7}
(\frac{\partial J}{\partial \Omega})_{\tilde{T}}=(\frac{\frac{\partial J}{\partial S}}{\frac{\partial \Omega}{\partial S}})_{\tilde{T}}=\frac{[(\frac{\partial ^2 \Omega}{\partial S\partial \tilde{T}})]_{\tilde{T}=\tilde{T}_c, S=S_c} (\tilde{T}-\tilde{T}_c)}{[(\frac{\partial J}{\partial S})]_{\tilde{T}=\tilde{T}_c, S=S_c}}.
\eea
Hence, the critical exponent $\gamma=1$. Also, by using Eqs. (\ref{crit4}) and (\ref{crit5}) we obtain the critical exponent $\delta$ defined at $\tilde{T}=\tilde{T}_c$ in the following form:
\bea
\label{crit7a}
\Omega(S,\tilde{T})-\Omega_c \propto(J-J_c)^3 \Rightarrow \ \delta= 3.
\eea
The results show that the calculated values of the critical exponents of MP black holes with quantum corrections at the critical point  correspond to a Van der Waals system and that they do not depend on the dimensionality of the system.
\section{THERMODYNAMIC GEOMETRY OF AN MP BLACK HOLE}
The phase transitions of black holes may also be viewed from the viewpoint  of thermodynamic state space (Ruppeiner)
geometry \cite{R1}. Thus, the Ruppeiner metric can be expressed as \cite{R2a}
   \bea
   dS^{2}=g_{ij}^{R} dX^{i}dX^{j},
   \eea
where $g_{ij}=-\frac{\partial ^{2} S(X^k)}{\partial X^i \partial X^j}$.
Also, the Weinhold metric can be defined as \cite{R2b}
   \bea
   dS^{2}_{W}=g_{ij}^{W} dX^{i}dX^{j}.
   \eea
Here, $g_{ij}^{W}=\frac{\partial ^{2} M(X^k)}{\partial X^i \partial X^j}$ and $i,j=1,2$, $X^{1}=J$, and $X^{2}=S$.
The relationship between the Ruppeiner metric and the Weinhold metric is expressed by
 \bea
   dS^{2}_{R}=\frac{1}{T} dS^{2}_{W}.
   \eea
   \begin{figure}[h]
\centering
\includegraphics[angle=0,width=7cm,keepaspectratio]{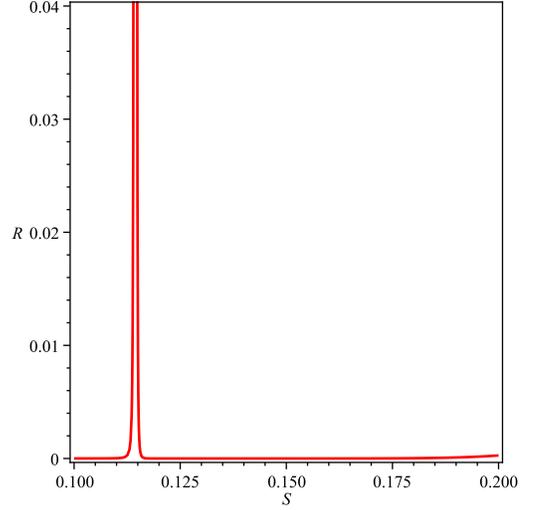}
\caption[]{{Ruppeiner curvature scalar ($R$) for $\Omega=\Omega _c$, $\beta _1=0.2$ and
$d=8$}.}
\label{figure 11}
\end{figure}
Here, $T$ is the temperature of the black hole system. The Ruppeiner curvature of MP black holes with quantum corrections is not flat at the critical point in the grand canonical ensemble and is divergent at $S=S_c$ (Fig. \ref{figure 11}). So, the second-order phase transition of MP black holes at the critical point can be investigated by using the Ruppeiner curvature formula in the grand canonical ensemble.\\
\indent If we consider the entropy of black holes as a function of the internal
energy $u = M - \Omega  J$ and $\Omega$ (angular velocity), then, for a fixed value of $J$, we may obtain
the Ruppeiner metric  and also the scalar curvature of geometry \cite{R2c}.
In this way, the first law of thermodynamics can be written as $du = TdS - Jd\Omega$. Since the thermodynamic metric introduced in Ruppeiner's theory is defined by the second derivatives of mass divided by temperature, we may replace the mass inducing the metric by the function regarded as the internal energy and the extensive variables in ordinary thermodynamic systems as follows:
  \bea
   g_{ij}^{R}=\frac{1}{T}\frac{\partial ^{2} M(X^k)}{\partial X^i \partial X^j}.
   \eea
Here, $i,j=1,2$, $X^{1}=\Omega$ and $X^{2}=u$. Using these, we find that the Ruppeiner curvature scalar $(R)$ is not flat but diverges at the critical point where the heat capacity is divergent in the canonical ensemble, indicating a second-order phase transition.
\section{CONCLUSION}
In this paper, we considered the thermodynamics of MP black holes in even dimensions
with $n=\frac{d-2}{2}$ nonzero equal spins $J$. We showed that while a second-order phase transition in the canonical ensemble is taking place in the semiclassical approximation, it is not possible to obtain a  critical behavior similar to that of a  Van der waals fluid. It was also found  that taking the corrections beyond the semiclassical approximation in the grand canonical ensemble gives rise to a critical behavior similar to a Van der Waals fluid. In this way, we obtain the value of the leading coefficient correction to be $\beta_{1}^{d=4}=\frac{1}{24\sqrt{3}}$ in order to find the universal number $\frac{PV}{T}=\frac{3}{8}$ in $d=4$. Also, the coefficient $\beta_1$ can be calculated for a given $d$ from the relation $\frac{P_c V_c}{T_c}=\frac{2d-5}{4d-8}$ \cite{m2}. It is straightforward to obtain  $\beta_{1}^{d=6}= \frac{7 (5-5^{\frac{1}{3}})}{1200}=0.01919$ and $\beta_{1}^{d=8}=0.01820$. These properties motivated us to check the validity of Ehrenfest's equations and the critical exponents. Our calculations showed that both Ehrenfest's equations are satisfied in the canonical and grand canonical ensemble, indicating that  a second-order phase transition happens. We also found that the corrected Gibbs free energy had positive values for any value of $\tilde{T}$ for $\Omega=\Omega_c$. We extended these calculations to the higher-order of correction terms. The results showed that the critical behavior and the critical exponents of black holes behave similar to a Van der Waals system.
In another part of this paper, we calculated the Ruppeiner curvature scalar $(R)$ and investigated
its behavior at the critical point. It was shown that R diverges exactly at the critical point where the specific heat and corrected specific heat are divergent in both the canonical and the grand canonical ensembles, respectively. This indicated that the Ruppeiner curvature formula could be exploited to investigate  the second-order phase transition for MP black holes at the critical point.

\section*{Acknowledgements}
This work has been supported financially by the Research Institute for Astronomy and Astrophysics of Maragha (RIAAM) under research project No. $1/2358$.


\begin{thebibliography}{9}

\bibitem{c1}
D. V. Fursaev, Phys. Rev. D {\bf 51,} R5352 (1995)
[arXiv:hep-th/9412161].

\bibitem{c2}
R. B. Mann and S. N. Solodukhin, Nucl. Phys.  {\bf B523,} 293
(1998) [arXiv:hep-th/9709064].

\bibitem{c3}
R. K. Kaul and P. Majumdar, Phys. Rev. Lett.  {\bf 84,} 5255 (2000)
[arXiv:gr-qc/0002040].

\bibitem{c4}
S. Carlip, Classical Quantum Gravity  {\bf 17,} 4175 (2000)
[arXiv:gr-qc/0005017];
M. R. Setare, Eur. Phys. J. C  {\bf 33,} (2004)
[arXiv:hep-th/0309134].



\bibitem{Haw}
S. W. Hawking, Commun. Math. Phys. {\bf 55,} 133 (1977).

\bibitem{RG}
K. Falls, D. F. Litim, [arXiv:1212.1821].

\bibitem{GUP}
M. Cavagliga, S. Das, Classical Quantum Gravity {\bf 21,} 4511 (2004)
[arXiv:hep-th/0404050].

\bibitem{GUP1}
K. Nozari and A. S. Sefiedgar, Phys. Lett. B. {\bf 635,} 156 (2006) [arXiv:gr-qc/0601116].



\bibitem{aman1}
J. E. Aman and N. Pidokrajt, Phys. Rev. D {\bf 73,} 024017 (2006)
[arXiv:hep-th/0510139].

\bibitem{aman2}
R. Emparan and H. S. Reall, Phys. Rev. Lett.  {\bf 88,} 101101
(2002) [arXiv:hep-th/0110260].



\bibitem{aman3}
J. E. Aman and N. Pidokrajt, [arXiv:1004.5550].


\bibitem{aman4}
R. Emparan and H. S. Reall, Living Rev. Relativity {\bf 11,} 6 (2008)
[arXiv:0801.3471].



\bibitem{aman5}
R. C. Myers, [arXiv:1111.1903].

\bibitem{aman6}
D. Astefanesei, M.J. Rodriguez, S.Theisen, J. High Energy Phys. 08 (2010) 046; D. Astefanesei, R. B. Mann, M. J. Rodriguez, 
C. Stelea, Classical Quantum Gravity {\bf 27}, 165004 (2010).





\bibitem{esp}
M. J. Rodriguez, in Proceedings of the Twelfth Marcel
Grossmann Meeting on General Relativity Paris, France
2009, edited by T. Damour, R. T. Jantzen, and R. Ruffini
(World Scientific, Singapore, 2012) [arXiv:1003.2411].


\bibitem{R1}
B. Mirza and M. Zamani-Nasab,  J. High Energy Phys. {\bf 06,} (2007) 059
[arXiv:0706.3450].

\bibitem{R2a}
G. Ruppeiner, [arXiv:0711.4328].

\bibitem{R2b}
F. Weinhold, J. Chem. Phys. {\bf 63,} 2479 (1975).

\bibitem{R2c}
J. Shen, R. G. Cai, B. Wang, R. K. Su, Int. J. Mod. Phys. A {\bf 22,} 11 (2007)
[arXiv:gr-qc/0512035].

\bibitem{R3}
R. G. Cai and J. H. Cho, Phys. Rev. D {\bf 60,} 067502 (1999)
[arXiv:hep-th/9803261].

\bibitem{R4}
H. Quevedo, Gen. Relativ. Gravit. {\bf 40,} 971 (2008)
[arXiv:0704.3102].

\bibitem{momen1}
A. Bravetti, D. Momeni, R. Myrzakulov, A. Altaibayeva, [arXiv:1303.2077]

\bibitem{E1}
S. Carlip and S. Vaidya, Classical Quantum Gravity {\bf 20,} 3827 (2003)
[arXiv:gr-qc/0306054].

\bibitem{E2}
Th. M. Nieuwenhuizen, Phys. Rev. Lett. {\bf 79,} 1317 (1997).

\bibitem{E3}
R. Banerjee, S. K. Modak and S. Samanta, Eur. Phys. J. C {\bf 70,} 317 (2010)
[arXiv:1002.0466].

\bibitem{wu1}
X. N. Wu, Phys. Rev. D {\bf 62}, 124023 (1999)

\bibitem{m1}
D. Kubiznak and R. B. Mann, J. High Energy Phys. {\bf 07} (2012)  033
[arXiv:1205.0559].


\bibitem{m2}
S. Gunasekaran and R. B. Mann, J. High Energy Phys. {\bf 11} (2012)  110  [arXiv:1208.6251].


\bibitem{m3}
C. Niu, Y. Tian and X. N. Wu, Phys. Rev. D {\bf 85,} 024017 (2012), [arXiv:1104.3066].

\bibitem{prig}
D. Kondepudi, I. Prigogine, {\it Modern Thermodynamics}  (John Wiley and Sons, NewYork, 1998).


\bibitem{crit1}
H. E. Stanley, {\it Introduction to phase transitions and critical phenomena}  (Oxford University Press, NewYork, 1987).


\bibitem{crit2}
R. Banerjee and D. Roychowdhuryz, Phys. Rev. D {\bf 85,} 044040 (2012)
[arXiv:1111.0147];
{\bf 85,} 104043 (2012)
[arXiv:1203.0118].

\bibitem{crit3}
Y. D. Tsai, X. N. Wu and Y. Yang, Phys. Rev. D {\bf 85,} 044005 (2012)
[arXiv:1104.0502].

\bibitem{E4}
R. Banerjee, S. K. Modak and S. Samanta, Phys. Rev. D {\bf 84,} 064024 (2011) [arXiv:hep-th/1005.4832].


\bibitem{E5}
R. Banerjee, S. Ghoshy and D. Roychowdhuryz, Phys. Lett. B {\bf 696,} 156 (2011)
[arXiv:1008.2644]; R. Banerjee, S. K. Modak and D. Roychowdhury, J. High Energy Phys. 10 (2012) 125 [arXiv:1106.3877].

\bibitem{E6}
R. Banerjee and D. Roychowdhury, J. High Energy Phys.  11  (2011)  004
[arXiv:1106.3877].

\bibitem{B1}
R. Banerjee and B. R. Majhi, J. High Energy Phys.  06 (2008) 095
[arXiv:0805.2220].

\bibitem{Yale}
A. Yale, Eur. Phys. J. C {\bf 71,}  1622 (2011)
 [arXiv:1102.5102].
\end{thebibliography}
\end{document}